\documentstyle[preprint,aps,epsfig]{revtex}
\begin{document}

\tighten
\draft
\preprint{
\vbox{
\hbox{April 1997}
\hbox{U.MD. PP\# 97-109}
\hbox{DOE/ER/40762-119}
}}

\title{Nucleon Strange Magnetic Moment and Relativistic Covariance}
\author{M. Malheiro$^{1,2}$ and W.Melnitchouk$^1$}
\address{$^1$	Department of Physics, 
		University of Maryland, 
		College Park, Maryland 20742-4111}
\address{$^2$	Instituto de F\'{\i}sica, 
		Universidade Federal Fluminense, 
		24210-340, Niter\'oi, R. J., Brazil}
\maketitle

\begin{abstract}
We calculate the corrections to the strange matrix elements of the
nucleon resulting from the breaking of rotational invariance on the
light-cone.
In the meson cloud model, the strange magnetic moment $\mu_S$
is seen to change sign once the spurious form factors arising
from this violation are subtracted.
The resulting $\mu_S$ is small and slightly positive, in agreement
with the trend of the recent data from the SAMPLE experiment.
The value of the strange magnetic form factor is predicted to be
largely $Q^2$ independent over the range accessible in upcoming
experiments.
\end{abstract}

Unraveling the strangeness content of the nucleon is one of the more
intriguing aspects of recent nucleon structure studies \cite{REP,BEISE}.
{}From the first polarized deep-inelastic muon-proton scattering
experiments at large $Q^2$ \cite{SPIN}, which suggested a negatively
polarized strange quark sea in the nucleon, and the studies of elastic 
neutrino-proton scattering cross sections at small $Q^2$ \cite{ELNU},
evidence has been accumulating for the non-trivial presence of strange
quarks in the nucleon.
Although the total number of strange and antistrange quarks in the nucleon
must of course be the same, their distributions in coordinate or momentum
space need not be equivalent \cite{JT}.
While perturbative QCD predicts equal $s$ and $\overline s$ distributions
in the nucleon, there is no fundamental symmetry which imposes this 
restriction outside of perturbation theory.
Understanding the strangeness properties of the nucleon should therefore
lead to a better appreciation of the role of non-perturbative QCD in
nucleon structure.

In response to the interest created by the original measurements
\cite{SPIN,ELNU} in the strangeness distributions in the nucleon,
several experiments involving parity-violating electron scattering
from protons and deuterons were proposed at MIT/Bates and Jefferson Lab
to measure neutral current form factors, from which various strange
quark matrix elements could be extracted.
Recently the SAMPLE Collaboration at MIT/Bates \cite{SAMPLE} reported 
the results of the first measurement of the strange magnetic form factor
of the proton, $G_M^S$, in the elastic scattering of 200 MeV electrons
from protons at backward angles with an average $Q^2 = 0.1$ GeV$^2$.
The trend of the data indicate a positive magnetic moment to the proton,
albeit with large errors \cite{SAMPLE}:
$G_M^S (Q^2 = 0.1$GeV$^2) = + 0.23 \pm 0.44$
nucleon magnetons (n.m.).

A non-zero strangeness content of the nucleon can be naturally accommodated
within a number of models of nucleon structure.
One of the simplest and most popular of these is the kaon cloud model.
Through quantum fluctuations, the nucleon here is viewed as having some
probability of dissociating into a virtual kaon and hyperon, both of
which carry strangeness quantum numbers.
Because of the very different masses and momentum distributions of
the kaon and hyperon, the overall strange and antistrange quark
distributions are therefore expected to be quite different, leading,
for example, to a non-zero strangeness radius, as well as strange
magnetic moment $\mu_S \equiv G_M^S(Q^2=0)$.

A common assumption in the application of the kaon cloud model is the
impulse approximation, in which one truncates the hadronic Fock space
at the one-meson level, and omits contributions arising from many-body
currents.
It is known, however, that the use of one-body currents alone for composite
systems leads to a violation of Lorentz covariance \cite{KEISTER}, which
would not be the case if one were to include the complete Fock space in
the calculation \cite{KF}.
In this note we will investigate in more detail the consequences of the
impulse approximation assumption, and in particular show that the magnitude
and even the sign of the strange magnetic moment are very sensitive to the
corrections which arise from the Lorentz covariance breaking.

One should note that the issue of relativistic covariance is relevant
both for light-front \cite{KF} as well as instant-form \cite{INSTANT,SJC}
quantization.
While on the light-front it is closely connected with the well known
problem of violation of rotational invariance, in the instant-form
approach the restriction to one-body currents also leads to a violation
of Lorentz covariance, as discussed in Ref.\cite{SJC}.
Irrespective of the orientation of the quantization surface, the problem
exists because one-body currents, which do not include interactions, do
not commute with the interaction-dependent generators of the Poincar\'e
group.
Consequently, an incorrect four-vector structure will appear in the matrix
elements of the current operator, resulting in the appearance in the full
electroweak current of additional unphysical, or spurious, form factors, 
which would not be present if the symmetry were preserved.
Nevertheless, it is still possible, to quantitatively estimate the extent
of the covariance violation within a specific model.

In the explicitly covariant formulation of light-front dynamics developed in
Refs.\cite{KF,KS}, a specific method was proposed to extract the nucleon's
physical form factors, excluding the spurious contributions.
It turns out that the unphysical form factors are most evident for
quantities that are small in magnitude, such as the neutron's
electromagnetic form factors, or the strangeness form factors.
The former was investigated in Ref.\cite{KF} in a simple constituent
quark model of the nucleon, where it was found that the difference between
the physical form factors and those obtained from the electromagnetic current 
without subtracting the spurious contributions can be as large as 50$\%$ at
high momentum transfers.
In view of the importance of determining accurately the strange properties
of the nucleon in current and upcoming experiments, one should obviously
re-examine the effect of the impulse approximation in the calculation of
strange form factors of the nucleon.
As discussed in Ref.\cite{MM}, there are a number of reasons why a light-cone
description of the dynamics is more attractive for this application.
We shall therefore restrict ourselves to the calculation of the form factors
within light-front quantization, and adopt the approach developed in 
Ref.\cite{KF} to expunge the unphysical contributions.

In the covariant light-front approach \cite{KF} the state vector is defined
on a light-front given by the invariant equation $n \cdot x = 0$, where $n$
is an {\em arbitrary} light-like four vector, $n^2 = 0$.
In constructing the most general form of the electroweak current on
the light-cone one has, in addition to the nucleon $p^\mu$ and current
$q^\mu$ four-momenta, the vector $n^\mu$, specifying the orientation
of the light-cone plane.
In principle, no physical quantity can depend on $n^\mu$.
Following Refs.\cite{KF,SJC}, one can show that the most general covariant 
form of the strange current operator $J_\mu^S$ can be written in terms of
5 form factors --- the usual $F_1^S$ and $F_2^S$, plus an additional three,
whose coefficients depend on $n^\mu$ \cite{KF}:
\begin{eqnarray}
\label{Jfull}
J_\mu^S
&=&
\overline u(p')
\left[ F_1^S(Q^2) \gamma_\mu\
    +\ F_2^S(Q^2) {i \sigma_{\mu\nu} q^\nu \over 2 M}
    +\ B_1^S(Q^2) \left( {\not\!n \over n \cdot p} 
		     - {1 \over (1+\eta) M} \right) P_\mu
\right.	\nonumber\\
& & \hspace*{2cm}
\left.
    +\ B_2^S(Q^2) {M \over n \cdot p} n_\mu
    +\ B_3^S(Q^2) {M^2 \over (n \cdot p)^2} \not\!n n_\mu
\right] u(p),
\end{eqnarray}
where $\eta = Q^2/4M^2$, $Q^2 = -q^2 = -(p' - p)^2$ and $P = p' + p$,
with the auxiliary condition $n \cdot q = 0$.
The new $n$-dependent structures appearing in the decomposition of
the amplitude for a spin 1/2 particle in Ref.\cite{KF} are essentially
the same as those in Ref.\cite{SJC}, differing only in the normalization
integral, which in \cite{KF} coincides with the electric form factor at
$Q^2=0$.
In principle, the form factors themselves can also depend on the vector
$n^\mu$, however, the $n$-dependence can only enter in the form of
the ratio $n \cdot p / n \cdot p'$ \cite{KF}, which is unity from the
condition $n \cdot q = 0$.

One can compare the covariant light-front formulation with the standard
approach of calculating matrix elements on the light-cone from the ``+'' 
component of the current \cite{LB}, $J_+^S = n \cdot J^S$, where a 
particular choice is made for $n$:\ $n^\mu = (1; 0, 0, -1)$,\
i.e. $t + z = 0$.
Taking the $\mu = +$ component of the current in Eq.(\ref{Jfull}) eliminates
two of the unphysical contributions, although one is still left with the
$B_1^S$ form factor. 
In fact, for the $\mu=+$ component Eq.(\ref{Jfull}) can be written:
\begin{mathletters}
\begin{eqnarray}
\label{J+}
J_+^S &=&
\overline u(p')
\left[ F_1^S(Q^2) \gamma_+\
    +\ F_2^S(Q^2) {i \sigma_{+ \nu} q^\nu \over 2 M}
    +\ 2 B_1^S(Q^2) \left( \gamma_+ - {p_+ \over (1+\eta) M} \right)
\right] u(p)		\\
\label{Jold}
&\equiv&
\overline u(p')
\left[ \widetilde{F}_1^S(Q^2) \gamma_+\
    +\ \widetilde{F}_2^S(Q^2) {i \sigma_{+ \nu} q^\nu \over 2 M}
\right] u(p),
\end{eqnarray}
\end{mathletters}%
where $\widetilde{F}_1^S = F_1^S + 2\eta B_1^S/(1+\eta)$ and
$\widetilde{F}_2^S = F_2^S + 2 B_1^S/(1+\eta)$
(henceforth the symbol $\widetilde{} $ refers to form factors
obtained from the ``+'' component of the current).
Note that even though Eq.(\ref{Jold}) takes a form just like that allowed
by Lorentz covariance constraints, the physical form factors are $F_{1,2}^S$,
and {\em not} $\widetilde{F}_{1,2}^S$.
The dependence of $\widetilde{F}_{1,2}^S$ on the spurious $B_1^S$ contribution 
is a manifestation of the violation of rotational invariance inherent in
the ``+'' prescription, which introduces an asymmetry between the $z$ and
$(x, y)$ axes.

An important observation from the above expression for $\widetilde{F}_1^S$
is that $\widetilde{F}_1^S(0) = F_1^S(0)$, which is essentially the splitting
function from which the deep-inelastic strange quark distribution in the meson
cloud model is calculated \cite{MM,DIS}.
Consequently, within the light-cone approach the structure functions can
be calculated unambiguously within the impulse approximation, without any 
unphysical $n$-dependent contributions.
Furthermore, for the measured Sachs electric and magnetic form factors
one can write the physical form factors as:
\begin{mathletters}
\begin{eqnarray}
\label{GE}
G_E^S &=& F_1^S - \eta F_2^S\ =\ \widetilde{G}_E^S\ ,		\\
\label{GM}
G_M^S &=& F_1^S + F_2^S\ =\ \widetilde{G}_M^S - 2 B_1^S\ .
\end{eqnarray}
\end{mathletters}%
Therefore, it is only the magnetic form factor which suffers contamination
from the unphysical contributions due to Lorentz covariance breaking,
while the electric form factor remains unchanged.
This can be understood from the fact that it is only the magnetic form
factor which is related to the matrix element of the spatial components
of the current, so that it receives contributions from the spurious
form factors associated with the rotational invariance breaking.

One should also note that, unlike for spin-1/2 and 1 particles, the matrix 
elements of the $J_+$ component for spin-0 particles such as the pion do not
involve unphysical form factors \cite{KF,SJC}.
For spin-1 systems, for example the deuteron, analogous expressions for the
form factors were obtained in Ref.\cite{KS}.
In fact, the problem of Lorentz covariance was explored for the deuteron
axial current using light-front dynamics in Ref.\cite{FHM}.
There the axial form factor was explicitly shown to be sensitive to the
choice of the matrix element of the axial current, which reflects the fact
that this operator must contain contributions from two-body currents if
rotational covariance is to be maintained \cite{FFS}.
This problem was recently reanalyzed by Keister \cite{KEISTER96} in the
manifestly covariant scheme of Gross \cite{GROSS}, and compared with
the light-front approach in an attempt to study the sensitivity to the
form of relativistic dynamics.

Having outlined the general approach to obtaining the physical form factors
of the nucleon consistently within the impulse approximation, we can now
proceed to investigate the effect on the strange magnetic form factor within
the kaon cloud model.
It is straightforward to evaluate the contributions to $B_1^S$ from the
interaction of the current with the kaon and hyperon components of the
nucleon.
(In practice, although the $\Sigma$, $\Sigma^*$, $\cdots$ hyperons can be
included, the $\Lambda$ is by far the most dominant contribution to the
strange nucleon form factors.)
One can define a generic strange form factor ${\cal G}^S$ as a sum of
$\Lambda$ and $K$ contributions:
\begin{eqnarray}
{\cal G}^S(Q^2)
&=& 
{\cal G}^{(\Lambda)}(Q^2) + {\cal G}^{(K)}(Q^2),
\end{eqnarray}
where ${\cal G}^S = G_M^S$ or $B_1^S$.
The results for the contributions to $\widetilde{G}_M^S$ were given
in Ref.\cite{MM}.
For the interaction with the $\Lambda$, the contribution to ${\cal G}^S$
can be written:
\begin{eqnarray}
\label{GMLam}
{\cal G}^{(\Lambda)}(Q^2)
&=&
Q_{\Lambda}{ g_{KN\Lambda}^2 \over 16 \pi^3 }
\int_0^1 dy \int { d^2{\bf k}_\perp \over y^2 (1-y) }
{ {\cal F}({\cal M}^2_{\Lambda K,i})\
  {\cal F}({\cal M}^2_{\Lambda K,f})
  \over ({\cal M}^2_{\Lambda K,i} - M^2)
	({\cal M}^2_{\Lambda K,f} - M^2) }\ 
  {\cal I}^{(\Lambda)},
\end{eqnarray}
where
\begin{equation}
\label{GB1Lam}
{\cal I}^{(\Lambda)}
=
\left\{
\begin{array}{l}
\Big( k_\perp^2 (1-2y) 
	- 2 ({\bf k}_\perp \cdot {\bf{\hat{q}}}_\perp)^2 (1-y)
	+ (M_{\Lambda}^2 - y^2 M^2)	
\\
\hspace*{4cm}
	+ (1-y)^2 q_\perp^2/4
\Big) / y  \hspace*{3cm} {\rm for}\ \ G_M^{(\Lambda)}\ ,	\\
{} \nonumber\\
\Big( - k_\perp^2 (1-3y)
	+ 2 ({\bf k}_\perp \cdot {\bf{\hat{q}}}_\perp)^2 (1-y) 
	- (1-y) (M_{\Lambda}^2 - y^2 M^2)
\\
\hspace*{4cm}
	- (1+y)(1-y)^2 q_\perp^2/4
\Big) / 2y\ \ \ \ \hspace*{1cm} {\rm for}\ \  B_1^{(\Lambda)}\ ,
\end{array}
\right.
\end{equation}
where ${\bf{\hat{q}}}_\perp^2 = {\bf q}_\perp^2 /q_\perp^2$ is a
unit vector.

For the $KN\Lambda$ vertex we assume a pseudoscalar $i\gamma_5$ 
interaction (the same results are obtained with a pseudovector coupling),
with $g_{KN\Lambda}$ the coupling constant, and 
${\cal F}$ the hadronic vertex function,
which we parameterize in terms of a cut-off mass $\Lambda_{K\Lambda}$
\cite{MM}:\
${\cal F}({\cal M}_{\Lambda K, i(f)}^2)
= ( \Lambda_{K\Lambda}^2 + M^2 )
/ ( \Lambda_{K\Lambda}^2 + {\cal M}^2_{\Lambda K, i(f)} )$.
The squared center of mass energies ${\cal M}^2_{\Lambda K, i(f)}$
in Eq.(\ref{GMLam}) depend on the invariant mass squared
${\cal M}^2\ =\ (p_K + p_\Lambda)^2
 \ =\ ({k_\perp^2 + M_{\Lambda}^2) / y}
    + ({k_\perp^2 + m_K^2) / (1-y)}$
of the $K\Lambda$ system,
where $m_K$ and $M_{\Lambda}$ are the masses of kaon and hyperon,
respectively, and $p_K$ and $p_\Lambda$ are their four-momenta.
For the sign of the strangeness we adopt the convention of Ref.\cite{JAF},
in which $Q_{\Lambda} = +1$ is the strangeness charge of the $\Lambda$.
One can check the consistency of Eqs.(\ref{GMLam}) and (\ref{GB1Lam})
using Eq.(\ref{GM}) to obtain the $\widetilde{G}_M^{(\Lambda)}$ as
calculated in Ref.\cite{MM}.

The kaon contribution to ${\cal G}^S$ can be written:
\begin{eqnarray}
\label{GMK}
{\cal G}^{(K)}(Q^2)
&=&
Q_{K}{ g_{KN\Lambda}^2 \over 16 \pi^3 }
\int_0^1 dy \int { d^2{\bf k}_\perp \over y (1-y)^2 }
{ {\cal F}({\cal M}^2_{K\Lambda,i})\
  {\cal F}({\cal M}^2_{K\Lambda,f})
  \over ({\cal M}^2_{K\Lambda,i} - M^2)
        ({\cal M}^2_{K\Lambda,f} - M^2) }\
  {\cal I}^{(K)},
\end{eqnarray}
where
\begin{equation}
\label{GBK}
{\cal I}^{(K)}
=
\left\{
\begin{array}{l}
2 k_\perp^2 - 2 ({\bf k}_\perp \cdot {\bf{\hat{q}}}_\perp)^2 
\hspace*{3cm} {\rm for}\ \ G_M^{(K)}\ ,   \\
{} \nonumber\\
\Big( k_\perp^2 (1-y)/y - 2 k_\perp^2
        + 2 ({\bf k}_\perp \cdot {\bf{\hat{q}}}_\perp)^2 
       + (M_{\Lambda}^2 - y^2 M^2)(1-y)/y
\\
\hspace*{2.45cm}
        - y (1-y) q_\perp^2/4
\Big) / 2 \ \ \ \hspace*{0.12cm} {\rm for}\ \  B_1^{(K)}\ ,
\end{array}
\right.
\end{equation}
and $Q_K = -1$ is the strangeness charge of the kaon.
The expression for $G_M^{(K)}$ especially is now very simple, depending
only on the relative orientation of the kaon and photon momenta.
In particular, there is no contribution to $G_M^{(K)}$ from the
configuration where ${\bf k}_\perp$ and ${\bf q}_\perp$ are parallel,
for which it would be impossible to flip the nucleon spin through the
interaction with the spin-0 kaon.
As we shall see below, the contribution from the $K$-interaction diagram
to the magnetic moment will be instrumental to its change of sign.

The results for the $B_1^S$ form factor turn out to be quite large in
magnitude, but opposite in sign compared with $\widetilde{G}_M^S$,
for the same value of the cut-off mass parameter $\Lambda_{K\Lambda}$.
The combined effect, illustrated in Fig.1 (solid curve), is a value
for $\mu_S$ which is almost independent of $\Lambda_{K\Lambda}$.
Also shown is the uncorrected value 
$\widetilde{\mu}_S \equiv \widetilde{G}_M^S(0)$ (dotted curve),
and $-2 B_1^S(0)$ (dashed curve).
In fact, the $B_1^{(\Lambda)}$ form factor is negative, making the 
contribution to the magnetic moment from the $\Lambda$-interaction 
more positive than for the uncorrected $\widetilde{G}_M^{(\Lambda)}$.
The same is true for the $K$-interaction contribution, where $B_1^{(K)}$,
which is also negative, gives a less negative contribution to $G_M^{(K)}$
compared with $\widetilde{G}_M^{K}$.
The final effect is that both the $K$-interaction and $\Lambda$-interaction
diagrams now contribute with almost the same magnitude and opposite sign,
making the total magnetic moment very small and positive: 
for $\Lambda_{K\Lambda} = 1$ GeV, one has
$G_M^{(\Lambda)}(0) = 0.044$ n.m. and $G_M^{(K)}(0) = -0.034$ n.m.,
for a total strange magnetic moment $\mu_S = + 0.010$ n.m.
This is to be compared with 
$\widetilde{G}_M^{(\Lambda)}(0) = 0.011$ n.m. and 
$\widetilde{G}_M^{(K)}(0) = -0.085$ n.m.,
with $\widetilde{\mu}_S = -0.074$ n.m.
In fact, all previous estimates of the strange magnetic moment in
meson cloud models have obtained negative values \cite{INSTANT,MM}.
The above results clearly illustrate the point that calculations
of form factors in the impulse approximation, which utilize only
single-particle operators, need to be treated with considerable care.
Even with the inclusion of so-called seagull terms \cite{SEAGULL},
which are examples of many-body contributions, the effects from
the Lorentz covariance breaking will still remain dominant.

In Fig.2 we show the total strange magnetic form factor $G_M^S(Q^2)$
as a function of $Q^2$, for values of $\Lambda_{K\Lambda}$ between
1 and 3 GeV (shaded region).
Surprisingly, the result seems almost entirely independent of $Q^2$,
which is a consequence of the cancellation between the $B_1^S$
and $\widetilde{G}_M^S$ terms above.
Also shown is the recent data point from the SAMPLE Collaboration,
which gave a value of 
$G_M^S (0.1$GeV$^2) = + 0.23 \pm 0.37 \pm 0.15 \pm 0.19$ n.m.,
where the errors are, respectively, statistical, systematic, and that
coming from the uncertainty due to the axial radiative corrections
and uncertainties in extracting $G_M^S$ from the measured asymmetry
\cite{SAMPLE}.

In summary, the breaking of Lorentz covariance due to the use of the
impulse approximation in the calculation of form factors of composite
systems manifests itself in the appearance of unphysical form factors
in the definition of the current.
We have investigated the consequences of this breaking within the meson
cloud model of the nucleon, by removing the spurious contributions
according to the prescription outlined in Ref.\cite{KF}.
For the strange magnetic form factor these contributions are significant,
and give rise to large cancellations, leading to an overall change of sign
for the physical $G_M^S$ form factor compared with earlier estimates
\cite{INSTANT,MM}.
The small (positive) value of the strange magnetic moment $\mu_S$ is
found to be largely independent of the details of the $KN\Lambda$ vertex
function, which is the main parameter in the model.
It will be of considerable interest to see to what extent the results
in Figs. 1 and 2 are supported in upcoming parity-violating experiments
at Jefferson Lab \cite{JEFF}, which will map out the $Q^2$ dependence
of the strange form factors over a larger range of kinematics.

\acknowledgements

We would like to thank X. Ji and J.-F. Mathiot for helpful discussions.
M.M. wishes to thank the TQHN group at the University of Maryland for
their hospitality during his extended visit, and the Brazilian agency
CAPES grant no. BEX1278/95-2 for the financial support which made this
visit possible.
This work was supported by the DOE grant DE-FG02-93ER-40762.


\begin{figure}
\label{fig1}
\epsfig{figure=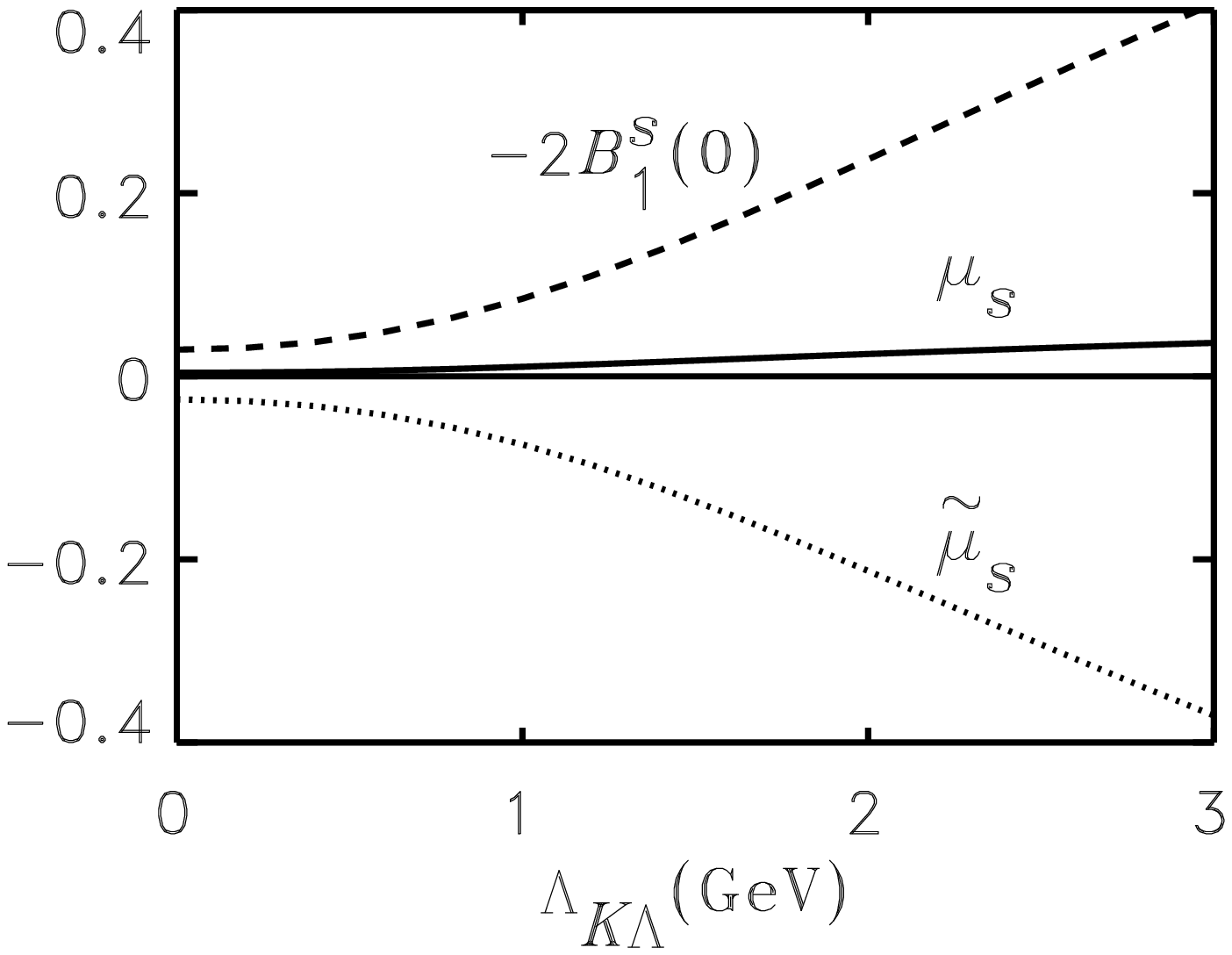,height=9cm}
\caption{Strange magnetic moment $\mu_S$ (solid) as a function of 
	meson--nucleon vertex function cut-off mass $\Lambda_{K\Lambda}$.
	Also shown are the unphysical $\widetilde{\mu}_S$ (dotted) and
	$-2 B_1^S(0)$ contributions (dashed).}
\end{figure}

\begin{figure}
\label{fig2}
\epsfig{figure=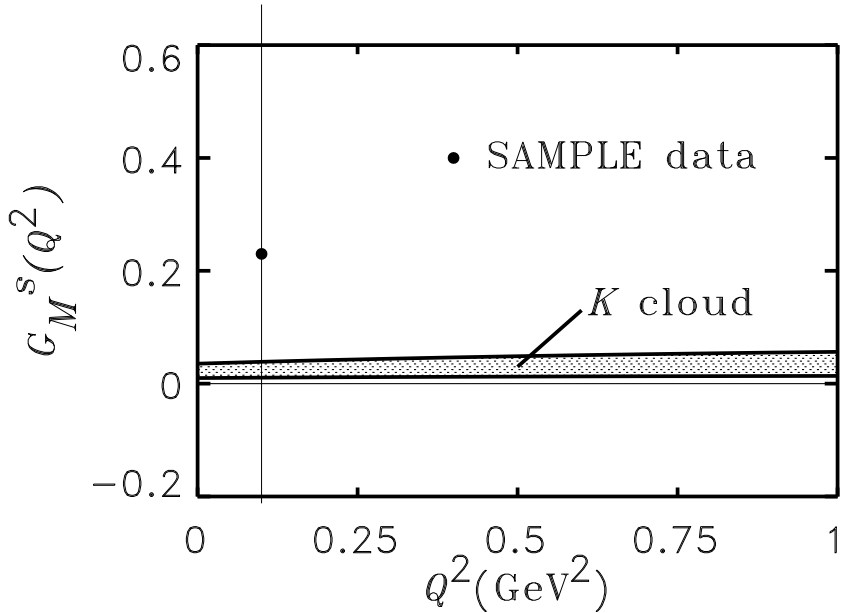,height=9cm}
\caption{Strange magnetic form factor of the proton $G_M^S(Q^2)$
	as a function of $Q^2$.
	The shaded region is the kaon cloud prediction, for 
	$\Lambda_{K\Lambda}$ = 1 (lower curve) and 3 (upper curve) GeV.
	The data point is from the SAMPLE experiment \protect\cite{SAMPLE}.}
\end{figure}

\end{document}